\documentclass[a4paper]{article}

\usepackage[english]{babel}
\usepackage[utf8x]{inputenc}
\usepackage{amsmath}
\usepackage{graphicx}
\usepackage[usenames,dvipsnames]{pstricks}
\usepackage{epsfig}
\usepackage{pst-grad} % For gradients
\usepackage{pst-plot} % For axes
\usepackage{epstopdf}

\usepackage{subcaption}
\title{A Complex Network Approach for Collaborative Recommendation}

\usepackage{babel}

\author{
  Ranveer Singh\\
  \text{Department of Mathematics}\\
  \text{IIT Jodhpur}  
      \and
      Bidyut Kumar Patra\\
      \text{Department of Computer Science and Engineering}\\
        \text{NIT Rourkela}
        \and
  Bibhas Adhikari\\
      \text{Department of Mathematics}\\
        \text{ IIT Kharagpur}
  }

\date{\today}

\begin{document}

\maketitle
\begin{center}
\textbf{Abstract}
\end{center}

Collaborative filtering (CF) is the most widely used and  successful approach for personalized 
service recommendations. Among the collaborative recommendation approaches,  neighborhood based approaches 
enjoy a huge amount of popularity, due to their simplicity, 
justifiability, efficiency and stability. Neighborhood based collaborative 
filtering approach finds  $K$  nearest neighbors   to  an active user or 
$K$ most similar rated items  to the target item for recommendation.  
Traditional similarity 
measures use ratings of co-rated items to find similarity 
between a pair of users. Therefore,  traditional similarity measures cannot 
compute  effective neighbors in sparse dataset. In this paper, we propose 
a two-phase  approach, which generates  user-user and  item-item networks using traditional 
similarity measures  in the first phase. In the second phase, 
two hybrid approaches \textbf{HB1}, \textbf{HB2}, which   utilize structural 
similarity of both the  network for finding $K$ nearest neighbors and $K$ most similar 
items to a target items are introduced. To show 
effectiveness of the measures, we compared performances of neighborhood based CFs using state-of-the-art 
similarity measures with our proposed structural similarity measures based CFs.  
Recommendation results on a set of real data show that proposed measures based CFs  
outperform existing measures based CFs in various evaluation metrics.

~\\

%\begin{keyword}
{\em Keyword-}
Collaborative filtering, neighborhood based CF, similarity measure, sparsity problem, 
structural similarity. 
%\end{keyword}

\section{Introduction}
In the era of information age, recommender systems (RS) has been established as 
an effective tool in various domains such as e-commerce, digital library, 
electronic media, on-line advertising, etc. % ~\cite{Resnick:1997, Billsus:2002, Linden:2003,   Miller:2003} 
\cite{b1}\cite{b2}\cite{b3}\cite{b4}. 
The recommender systems provide personalized suggestions about  products 
or services  to individual user filtering through large product or item space. 
The most successful and widely accepted recommendation technique 
is collaborative filtering (CF), which  utilizes only user-item interaction for 
providing recommendation unlike content based approach \cite{b5} \cite{b6}. % ~\cite{Lang95, Pazzani:1997}. 

The  CF technique is based on the intuition  that  users who has expressed similar interest 
earlier will have  alike choice in future also. The approaches in CF can be classified into 
two major categories, {\em viz. } {\em model based CF} and {\em neighborhood based CF}.

Model-based CF algorithms learn a model from the training data and  
subsequently, the model is utilized  for recommendations \cite{b9}\cite{b10}\cite{b11}. 
Main advantage of the model-based approach is that it does not need to 
access whole rating data once model is built. Few model based approaches provide more 
accurate results than neighborhood based CF \cite{b12}\cite{b13}. 
However, most of the electronic retailers such as Amazon, Netflix deployed
neighborhood based recommender systems to help out their customers. This is due to the 
fact that neighborhood based approach is simple, intuitive and it does not have learning 
phase so it can provide immediate response to new user after receiving upon her feedback. 
Neighborhood based collaborative algorithms are further classified into two categories, 
{\em viz. } {\em user based } and {\em item based CF}. 

The user based CF is based on a principle  that an item might be interesting to an active 
user in future if a set of neighbors of the active user have appreciated the item. In item 
based CF,  an item is recommended to an active user if she has appreciated similar items in past. 
$Neighborhood$ $based$ $CF$ recommendation extracts user-user 
or item-item similarity generally using Pearson Correlation Coefficient(PCC)and its variants, 
slope-one predictor from user-item matrix to predict rating of user for new item \cite{t1}. 
Item based CF  is preferred over user based CF if number of items is smaller in number 
compared to the number of  users in the system. 

Generally, neighborhood based CF uses a similarity measure for finding 
neighbors of an active user or finding similar items to the candidate item. 
Traditional similarity measures such as pearson correlation coefficient, 
cosine similarity and their variants are frequently used for 
computing similarity between a pair of users or between a pair 
of items  \cite{b15}. 
In these measures,  similarity between a pair of users is computed based on the ratings made by both users 
on the common items (co-rated item). Likewise, item similarity is  computed using the ratings provided by users 
who rated both the items.  
However, correlation based measures perform poorly if there are no sufficient numbers of co-rated items 
in a given rating data. Therefore, correlation based measure and its variants are not suitable 
in a sparse data in which number of ratings by individual user is less and number of  co-rated items is 
few or none \cite{t11}.

In this paper, we  propose a novel approach for computing similarity between a 
pair of users (items) in sparse data. In the proposed approach,  
a user-user (item-item) network is generated using 
pearson correlation (adjusted cosine) for computing similarity 
between a pair of users (items). 
Having  generated the networks, we exploit the structures of the network  
for addressing few drawbacks of neighborhood based CF. Having generated the
networks, we exploit the structures of the network for computing similarity in sparse data 
and predictions for an item which receives ratings from few users.  The approach is tested 
on real rating datasets. The contributions in this paper are  summarized as follow. 

\begin{itemize}
 \item We propose a novel approach to utilize traditional similarity measures to generate 
 user-user and item-item networks. The generated networks help in establishing transitive  
  interaction between a pair of users, which is difficult to capture using traditional measures. 
  
\item The user based CF fails to predict rating of an item if it receives rating from few users (less than 
     $K$ users). Structure of the networks are exploited to address this problem by combining item based CF 
     with the user based CF. Two algorithms termed as HB1 and HB2 are introduced for this purpose. 
     
\item We discuss the drawback of the use of F1 measure in recommendation scenario and 
     introduce a new metric to evaluate qualitative performance of the collaborative filtering algorithms  to 
     capture the variation of rating  provided by individual user. 
     
\item To show the effectiveness of the proposed approach in sparse dataset, we implemented neighborhood based 
      CF using traditional similarity measures and neighborhood based CF using proposed approach. 
     
\end{itemize}

The rest of the paper is organized as follows. The background and related 
research are discussed in Section \ref{sec:back}. The proposed novel 
approach is introduced in Section \ref{sec:prop}. Experimental results and evaluation of the 
proposed approach are reported in Section \ref{sec:result}. We conclude the paper in Section  
\ref{sec:con}.

\section{Background and related work}\label{sec:back}
In this section, we discuss working principle of neighborhood based approach in 
detail and different similarity measures introduced in literature in-order to 
increase performance of recommendation systems over past decades.
\subsection{Neighborhood-Based Approach}

The neighborhood or memory based approach is introduced in the 
GroupLens Usenet article\cite{b18} recommender and has gained popularity 
due to its wide application in commercial domain \cite{b3}\cite{b14}\cite{b19}. 
This approach uses the entire rating dataset to generate a prediction for an item 
(product) or a list of recommended items for an active user. Let $R = (r_{ui})^{M×N}$ 
be a given rating matrix (dataset) in a CF based recommender system, 
where each entry $r_{ui}$ represents a rating value made by $u^{th}$ 
user $U_{u}$ on $i^{th}$ item $I_{i}$. Generally, rating values are 
integers within a rating domain(RD), e.g 1-5 in MovieLens dataset. 
An entry $r_{ui}=0$ indicates user $U_{u}$ has not rated the item $I_{i}$. 
The prediction task of neighborhood-based CF algorithm is to predict rating 
of the  $i^{th}$ item either using the neighborhood information of $u^{th}$ 
user (user-based method) or using neighborhood information of $i^{th}$  
item (item-based method).

Neighborhood-based Prediction method can be divided into two parts 
User-based and Item-based. User based methods predicts based upon on 
ratings of $i^{th}$ item made by the neighbors of the $u_{th}$ 
user \cite{b15}\cite{b20}. This method will computes similarity of 
the active user $U_{u}$ to other users $U_{p}$,$p=1,2...M, p\neq u$. 
Then $K$ closest users are selected to form neighborhood of the active 
user. Finally, it predicts a rating $\hat{r}_{ui} $ of the $i^{th}$ item 
using the following equation.
\begin{equation}
\hat{r}_{ui}=\bar{r}_{u}+\frac{\sum_{k=1}^{K}s(U_{u},U_{k})(r_{ki}-\bar{r}_{k})}{\sum_{k=1}^{K}|s(U_{u},U_{k})|}
\end{equation} 

where, $\bar{r}_{u}$ is the average of the ratings made by 
user $U_{u}$. $s(U_{u},U_{k})$ denotes similarity value between
user $U_{u}$ and its $k^{th}$ neighbor. $\bar{r}_{k}$ is the average of  
ratings made by $k^{th}$ neighbor of the user $U_{u}$, and $r_{ki}$ is 
the rating made by $k^{th}$ neighbor on $i^{th}$ item.
         
Item-based collaborative filtering \cite{b3} has been 
deployed by world’s largest online retailer Amazon Inc \cite{b3}. 
It computes similarity between target item $I_{i}$ and all other 
items $I_{j}$ , $j =1,...N i\ne j$ to find $K$ most similar items. 
Finally, unknown rating $\hat{r}_{ui}$ is predicted using the ratings 
on these $K$ items made by the active user $U_{u}$.

\begin{equation}
\hat{r}_{ui}=\bar{r}_{i}+\frac{\sum_{k=1}^{K}s(I_{i},I_{k})(r_{uk}-\bar{r}_{k})}{\sum_{k=1}^{K}|s(I_{i},I_{k})|}
\end{equation}

where, $\bar{r}_{i}$ is the average of the ratings made by 
all users on items $I_{i}, s(I_{i},I_{k})$
denotes the similarity between the target item $I_{i}$ and 
the $k^{th}$ similar item, and $r_{uk}$ is the rating made by 
the active user on the $k^{th}$ similar item of $I_{i}$.

Similarity computation is a vital step in the neighborhood based 
collaborative filtering. Many similarity measures have been introduced 
in various domains such as machine learning, information retrieval, 
statistics, etc. Researchers and
practitioners in recommender system community used them directly or 
invented new similarity measure to suit the purpose. We discuss them briefly next.

\subsection{Similarity Measures in CF}
 Traditional measures such as pearson correlation coefficient (PC), 
 cosine  similarity are frequently used in recommendation systems. 
 The cosine similarity is very popular measure in information retrieval 
 domain. To compute similarity between two users $U$ and $V$ , they are 
 considered as the
 two rating vectors of $n$ dimensions, i.e., $U, V \in N^{n}_{0}$
 , where $N_{0}$ is the set of natural numbers including 0. Then, 
 similarity value between two users is the cosine of the angle 
 between $U$ and $V$. Cosine similarity is popular in item based CF. 
 However, cosine similarity does not consider the different rating 
 scales (ranges) provided by the individual user while computing 
 similarity between a pair of items. Adjusted cosine similarity 
 measure addresses this drawback by subtracting the corresponding 
 user average from the rating of the item. It computes linear 
 correlation between ratings of the two items.\\
 Pearson correlation coefficient (PCC) is very popular measure 
 in user-based  collaborative filtering. The PCC measures how two users 
 (items) are linearly  related to each other. Having identified co-rated 
 items between users $U$ and  $V$ , PCC computes correlation between 
 them \cite{b26}. The value of PCC ranges in [−1 +1], The value +1 indicates 
 highly correlated and −1 indicates negatively co-related to each other. 
 Likewise, similarity between two items $I$ and $J$ can also be computed using PCC. 
 Constrained Pearson  correlation coefficient (CPCC) is a variant of 
 PCC in which an absolute reference (median in the rating scale) is used 
 instead of corresponding user’s rating average. Jaccard only considers the 
 number of common ratings between two users. The basic idea is that users are 
 more similar if they have more common ratings. Though profoundly used PCC and 
 its variants suffer from some serious  drawbacks described in section 4.

PIP is the most popular (cited) measure after traditional similarity measures 
in RS. The PIP measure captures three important aspects (factors) namely, proximity, 
impact and popularity between a pair of ratings on the same item \cite{b26}.
The proximity factor is the simple arithmetic difference between two ratings on 
an item with an option of imposing penalty if they disagree in ratings. 
The agreement (disagreement) is decided with respect to an absolute reference, 
i.e., median of the rating scale. The impact factor shows how strongly an item 
is preferred or disliked by users. It imposes penalty if ratings are not in the 
same side of the median. Popularity factor gives important to a rating which is 
far away from the item’s average rating. This factor captures global information 
of the concerned item. The PIP computes these three factors between each pair of
co-rated items. PIP based CF outperforms correlation based CF in providing 
recommendations to the new users.
  
Haifeng Liu et al. introduced a new similarity measure called NHSM 
(new heuristic similarity model), which addresses the drawbacks of 
PIP based measure recently. They put an argument that PIP based measure 
unnecessarily penalizes more than once while computing proximity and impact 
factors. They adopted a non-linear function for computing three factors, 
namely, proximity,significance, singularity in the same line of PIP based 
measure. Finally, these factors are combined with modified Jaccard similarity measure \cite{b33}.

Koen Verstrepen and Bart Goethals proposed a method that unifies user- 
and item based similarity algorithms \cite{t5}, but it is suitable for binary scale. 
  
 Before proposing our method, let us first analysis the problem with 
 methods which use co-rated items through experiments on different datasets.  These problems motivated us to formulate our proposed methods.

\section{Proposed method for rating prediction}\label{sec:prop}

As mentioned earlier, we employ the well developed concept of similarities of nodes in a network for the prediction of entries of a rating matrix when a network is generated by using the given data. We propose to generate an user-user (resp. item-item) weighted network with node set as the set of users (resp. items) and the links in the network are defined by the PCC similarity of the users (resp. items) in the given data. Once the network is generated, a local similarity metric for nodes reveals insights about the connectivity structure of neighbors of a given node, and a global similarity metric provides the understanding of how a node is correlated with rest of the nodes in the network. Thus, a network approach to determine correlations between users or items provides a holistic outlook into the interpretation of a data.

Further, in order to reduce sparsity in the data for rating prediction, we introduce the concept of intermediate rating for an item by an user who has not rated the corresponding item. Thus, we use both the user-user and item-item network structural similarities to propose new metrics for rating prediction.

 \subsection{Network generation and structural similarities}

Let $\mathcal{U}$ and $\mathcal{I}$ denote the set of users and items respectively of a given data set. The adjacency matrix $A=[A_{ij}]$ for the user-user (resp. item-item) weighted network is defined by $A_{i,j}=\mbox{PCC}(i,j)$ where $i, j$ denote users (resp. items) and PCC$(i,j)$ denotes the PCC similarity between $i$th and $j$th nodes. Thus, in the user-user (resp. item-item) network, the users (resp. items) are represented by nodes and links between any two nodes are assigned with weight $\mbox{PCC}(i,j)$ if $\mbox{PCC}(i,j)\neq 0$ otherwise the nodes are not linked. The size of the user-user (resp. item-item) network is given by $|\mathcal{U}|$ (resp. $|\mathcal{I}|$) where $|X|$ denotes the carnality of a set $X.$

In this paper, we consider the following structural similarities for the user-user or item-item network. Let $A$ denote the adjacency matrix associated a network $G$. For a node $i$ of $G$, let $\gamma(i)$ denote the set of neighbors of $i$.  \begin{itemize} \item Common Neighbors (CN)\cite{t4}. The CN similarity of two distinct nodes $i$ and $j$ is defined by \begin{equation*}
s^{CN}_{ij}=|\gamma(i) \cap \gamma(j)|.
\end{equation*}
It is obvious that $s^{CN}_{ij} = [A^{2}]_{ij},$ the number of different paths with length $2$ connecting $i$ and $j$. So, more the number of common neighbors between two nodes, more is the value $s^{CN}_{ij}$ between them.

\item Jaccard Similarity \cite{t4}: This index was proposed by Jaccard over a hundred years ago, and is defined as
\begin{equation*}
s^{\mbox{Jaccard}}_{ij}=\frac{|\gamma(i) \cap \gamma(j)|}{{|\gamma(i) \cup \gamma(j)|}}
\end{equation*} for any two distinct nodes $i$ and $j$ in the network.

\item Katz Similarity \cite{t4}. The Katz similarity of two distinct nodes $i, j$ is defined by
\begin{equation}
s^{Katz}_{ij}= \sum_{l=1}^{\infty} \beta^{l}\cdot|\mbox{paths}<l>_{ij}|=\beta A_{ij} + \beta ^{2}[A^{2}]_{ij}+\beta ^{3}[A^{3}]_{ij}+\hdots,
\end{equation}
where paths$<l>_{ij}$  denotes the set of all paths with length $l$
connecting $i$ and $j$, $[A^p]_{ij}$ denotes the $ij$th entry of the matrix $A^p, p$ is a positive integer and $0< \beta $ is a free parameter (i.e.,
the damping factor) controlling the path weights. Obviously, a very
small $\beta$ yields a measurement close to $CN$, because the long
paths contribute very little. The Katz similarity matrix can be written as
\begin{equation*}
S^{Katz}= (I-\beta A)^{-1} - I
\end{equation*}
where $\beta$ is less than the reciprocal of the largest
eigenvalue of matrix $A.$ Thus, Katz similarity of two nodes is more if the number of paths of shorter length between them is more.\\
Let $\lambda_{1}$ be largest eigenvalue of $A$ in magnitude. We have set $\beta= \frac{0.85}{\lambda_{1}}$. This value of $\beta$ is used as damping factor of Google’s Page-Rank sorting algorithm \cite{b33}.
\end{itemize}

Using the above mentioned similarity indices for users and items, one could predict the rating $r_{ui}$ of $I_i$th item by user $U_u$ as $\widehat{r}_{ui}$ by using the formulae (1) and (2).

%We can make a User-User or Item-Item network using different similarities.
%In present work we have used PCC similarity. In network two users $i$ and
%$j$ are connected if similarity is greater than certain thrashold PCC. So
%we will have a Adjacency matrix $A$ of order $|U|$ (Total no. of users)or
%$|I|$ (Total no. of items). $A(i,j)=1$, if $i$ and $j$ are connected else
%$A(i,j)=0$. Adjacency matrix $A$ can be weighted also, with weights equal
%to similarity i.e $A(i,j)=Sim^{pcc}(i,j)$ if $i$ and $j$ are connected
%else $A(i,j)=0$. In the same way we can generate Item-Item Network. For
%our present work we have taken weighted connections, threshold for
%experiment 1 is 0.65 and for experiment 2 is 0.40.   \\
%
%What benefit a network will do ?. Once we get a network we can use
%structural similarities between any two users or items. Structural
%similarities are based solely on the network structure. These similarities
%are always a real number and we can get enough similar user or items(neighbors)
%for the prediction. So sparsity get vanished as well as we will
%have good chance of having sufficient numbers of neighbors for the prediction.
%
%
% We can divide structural similarities mainly into two parts
% depending upon range of network topology they use (1). Local
% Similarities (2). Global Similarities \cite{t4}.

 \begin{table}[htp]
  \begin{center}
  \begin{tabular}{|c|c|c|c|c|c|c|c|c|} \hline\hline
    \emph{Dataset } & \emph{Purpose} & $|\mathcal{U}|$ & $|\mathcal{I}|$ &\emph{\#Ratings} & $|\mathcal{U}|/|\mathcal{I}|$ & $\kappa$ & \emph{RD} \\
    \hline \hline

   $Movielens$  & Movie& 6040 & 3706& 1 M& 1.6298 & 4.46 & [1-5] \\
   $Yahoo$  & Music & 15400 & 1000 & 0.3 M & 15.4 & 2.024& [1-5] \\
   $Netflix$  & Movie& 4141 & 9318& 1M & 0.4444 & 2.64& [1-5] \\

  \hline\hline
  \end{tabular}
  \end{center}
  \caption{Description of the datasets used in the experiments.} \label{t1}
  \end{table}

\subsection{Network based hybrid approach for rating prediction}

In this section, we introduce two new methods for rating prediction using both user-user and item-item networks constructed by the given data as mentioned above. Note that, in spite of significant advancement of calculation of user-user or item-item similarity in the network approach, the curse of sparsity hinders a finer prediction of ratings. For instance, if only a few users rated a particular item, network similarity of users suffer from accuracy keeping in mind that we can only use similarity of users who have rated a particular item $I_{i}$ to which rating is to be predicted for an user $U_{u}$. For the prediction of $r_{ui},$ the rating of user $U_u$ for the item $I_i$, the $K$-neighbors problem \cite{t11} is concerned with the existence of minimum $K$ number users who rated the item $i$. Thus, in order to get rid of $K$-neighbors problem, we introduce the idea of intermediate rating ($\mbox{IR}_{k}$) as follows.

Let $N_u$ be the number of users who have rated an item $I_i$ and $N_u< K.$ Then, at first, we determine $K - N_{u}$ users $U^{s}_{1},U^{s}_{2}, \hdots, U^{s}_{K-N_{u}}$ best similar to $U_{u}$. For these users, we predict rating of user $U^{s}_{k}, k=1:K-N_u$ for the item $I_{i}$ using item based similarity as
\begin{equation}\label{IRk}
\mbox{IR}_{k}=\bar{r}_{i}+\frac{\sum_{j=1}^{K^{I}}s(I_{i},I_{j})(r_{U^{s}_{k}j}-\bar{r}_{j})}{\sum_{j=1}^{K^{I}}|s(I_{i},I_{j})|}
\end{equation}
where $K^{I}$ is number of items whose similarities we have to use
for prediction of $\mbox{IR}_{k}$ and $s(I_{i},I_{j})=s^{\mbox{Jaccard}}_{ij}$. For our present work we have set $K^{I}=10$,
but if similar user $U^{s}_{k}$ has rated less than 10 items we have
used similarity of that many items for intermediate prediction. 

Nevertheless, the $K$-neighbors problem can also be avoided for prediction of $r_{ui}$ by selecting best $K$ users similar to $U_u$ without considering whether they have rated the item $I_i.$ If they have rated the item $I_i$ then we use those ratings for prediction, else the value of $\mbox{IR}_k$ can be used for the same.

Now, we propose the following formulae for prediction of $r_{ui}.$ \begin{itemize} \item HB1: \begin{equation}
\hat{r}_{ui}= \begin{cases} \bar{r}_{u}+\frac{\sum_{k=1}^{K}s(U_{u},U_{k})(r_{ki}-\bar{r}_{k})}{\sum_{k=1}^{K}|s(U_{u},U_{k})|} &\mbox{if } N_{u}\ge K\\ \\
\bar{r}_{u}+\frac{\sum_{k=1}^{N_{u}}s(U_{u},U_{k})(r_{ki}-\bar{r}_{k})}{\sum_{k=1}^{N_{u}}|s(U_{u},U_{k})|}+\frac{\sum_{k=1}^{K-N_{u}}s(U_{u},U^{s}_{k})(\mbox{IR}_{k}-\bar{r}_{k})}{\sum_{k=1}^{K-N_{u}}|s(U_{u},U^{s}_{k})|} &\mbox{if } N_{u}< K
\end{cases}
\end{equation}

\item HB2: \begin{equation}
\hat{r}_{ui}= \begin{cases} \bar{r}_{u}+\frac{\sum_{k=1}^{K}s(U_{u},U_{k})(r_{ki}-\bar{r}_{k})}{\sum_{k=1}^{K}|s(U_{u},U_{k})|} &\mbox{if } r_{ki}\ne 0\\ \\
\bar{r}_{u}+\frac{\sum_{k=1}^{K}s(U_{u},U_{k})(\mbox{IR}_{k}-\bar{r}_{k})}{\sum_{k=1}^{K}|s(U_{u},U_{k})|} &\mbox{if } r_{ki}= 0
\end{cases}
\end{equation}\end{itemize} where the similarity between the users are calculated by the Jaccard similarity in user-user network.

  \begin{center}  
  \begin{figure}
  
  \begin{subfigure}{.32\textwidth}
     \includegraphics[width=4.1cm]{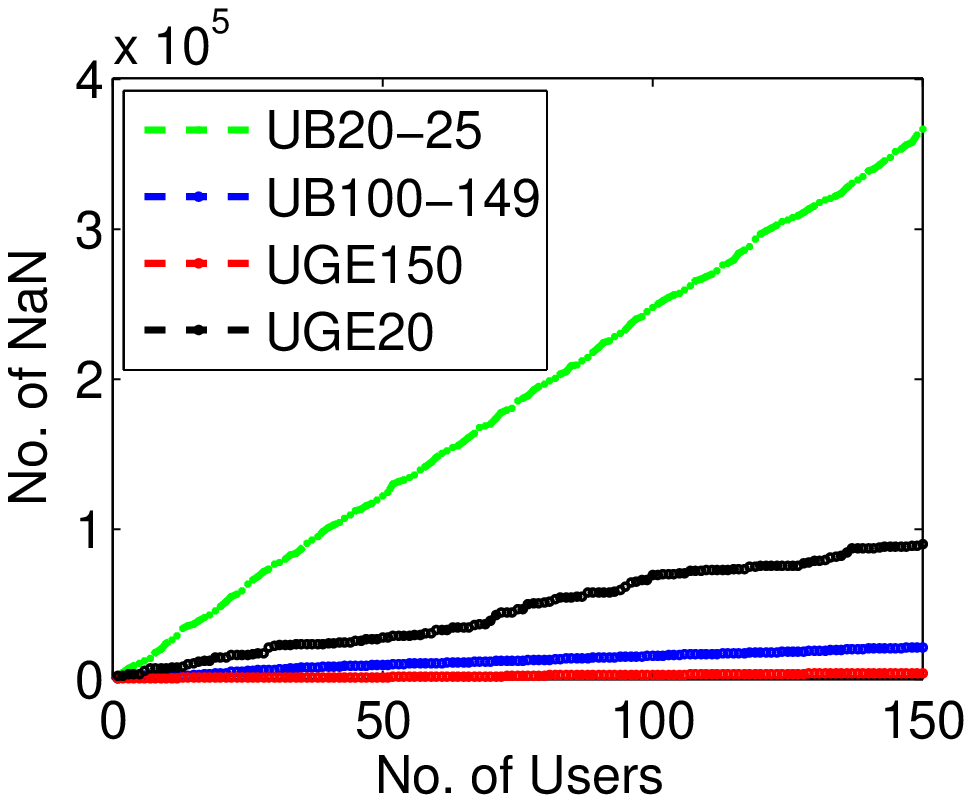}
    \caption{}
    \label{fig:sub41}
  \end{subfigure}%
  \begin{subfigure}{.3\textwidth}
     \includegraphics[width=4.1cm]{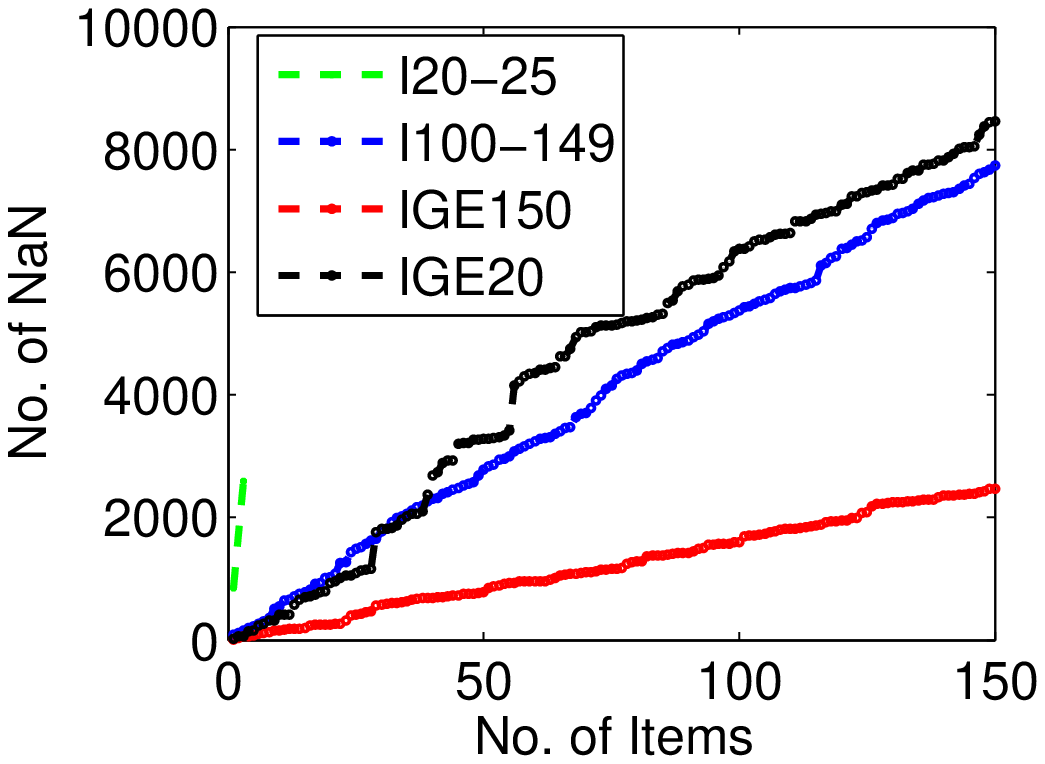}
    \caption{}
    \label{fig:sub42}
  \end{subfigure} 
  %\begin{center}
  \begin{subfigure}{.3\textwidth}
     \includegraphics[width=4.1cm]{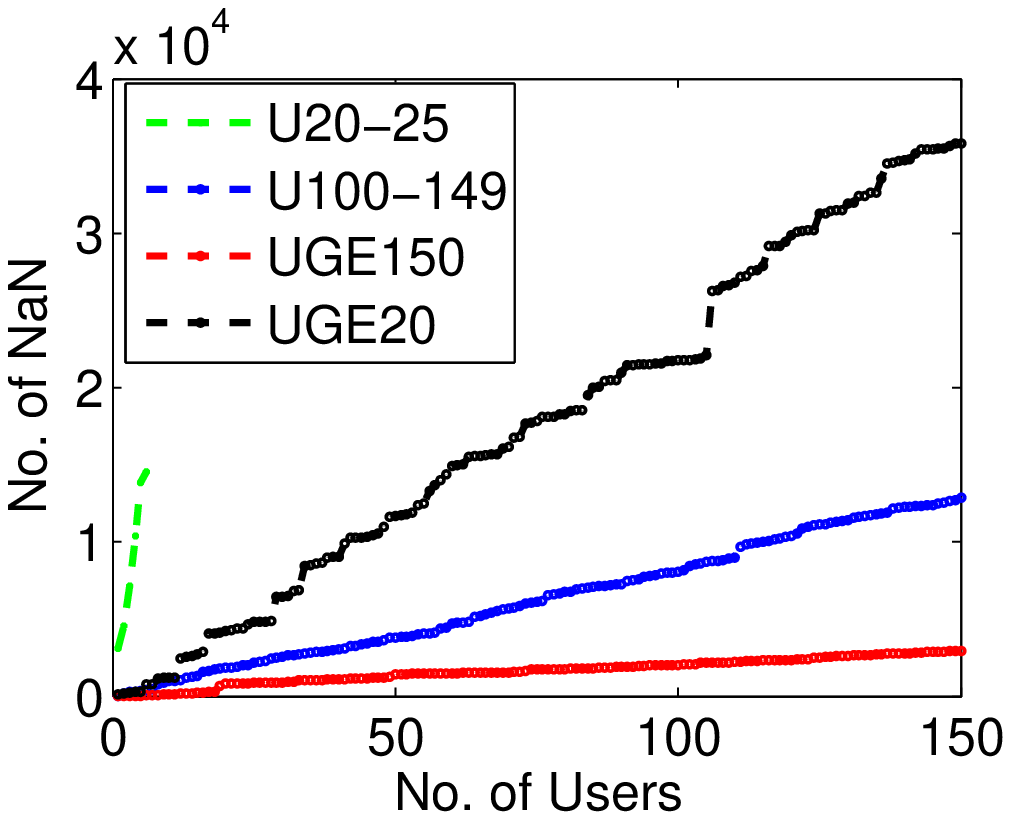}
    \caption{}
    \label{fig:sub42}
  \end{subfigure}
  %\end{center}
  \caption{ No. of NaNs (a)ML (b)YH (c)NF }
  \label{f1}
  
  \end{figure}
  
  \end{center}

  \section{Experimental Evaluation}\label{sec:result}
  
  We conducted  experiments on three real datasets, namely,  Movielens (ML), 
  Yahoo (YH) and Netflix (NF). Detailed description is given in 
  Table \ref{t1}. We utilized these datasets in two different ways to 
  show the efficiency of our approaches on original dataset (Experiment 1 setup) as well as in  sparse 
  scenarios (Experiment 2 setup).  
  We apply user-based method on Movielens and Netflix datasets and item-based method on  Yahoo dataset as 
  items are quite less compared to users in YH dataset.
  
  \subsection{Experiment 1 Setup}
   This setup    is to show how the predictions differ for different users (items) based on the 
   numbers of ratings a user made (an item received). We  divided each dataset 
   into 4 parts as described in  (Table \ref{t3}).

  \begin{table}[t]
   \begin{center}
   \begin{tabular}{|c|c|c|c|c|} \hline 
   \emph{Dataset}&\multicolumn{4}{|c|}{Ratings per user (item)} \\ \cline{2-5} 

 &  20-25 & 100-149 & $\geq$150 &$\geq $ 20 \\  \hline
    $Movielens$  & 491 & 849  & 2096 & 6040  \\ \hline 
    $Yahoo$  &  (3)& (243) & (518) & (998) \\ \hline
    $Netflix$  & 6 & 551 & 1991& 4139 \\ 
   \hline
   \end{tabular}
   \end{center}
   \caption{Division of datasets used in experiment 1 setup. } \label{t3}
   \end{table}

  \begin{enumerate}
  \item \textbf{U20-25 ( I20-25)}:  The \textbf{U20-25} is a group  of users 
  who have rated  number of items between 20-25. There are  total 491  users 
  in ML dataset. Likewise, \textbf{I20-25} is set of items which are rated by  
  number of users between 20-25. There are only three (3) items in YH. So,  this set  consists 
  of very sparse user (item) vectors. We see the effect of data sparsity on 
  prediction for these users (items). Neighborhood selection is done from whole dataset.  
  
  \item \textbf{U100-149 (I100-149)}:  \textbf{U100-149} is the set of 
  users who have rated a significant number of  items in ML and Netflix datasets. 
  Similarily, \textbf{I100-149} is set of items which are rated by 
    number of users between 100 and 149. There are such 243  items in YH. So,  this set  
  consists of users (items) which have significant ratings.
  
  \item \textbf{UGE150 (IGE150)} :  \textbf{UGE150} is set of users who 
  have rated more than 149 number of items.  
  Similarily, \textbf{IGE150} is set of items which are rated by more than 149 number 
  of users. 
  
  \item \textbf{UGE20 (IGE20)}:  This set consists of 
user (item) vectors from  whole dataset.

  \end{enumerate} 
  
  To compare prediction performance for different category we  randomly select 
  150 users (items) from each category. If total number  of users (items) in a set is less than 
  150, we select all of them in that category. For each  user,  we  randomly delete 15 ratings. 
  Deleting more than 15 entries for test purpose will mostly result in void 
  or very sparse vector(only 1 or 2 ratings). We predict these deleted ratings using state of art 
  methods and methods proposed in this paper.    
  
 \subsection{Experiment Setup 2}
 Main objective of the experiment setup 2 is to show the performance of our network based 
 similarity measures on sparse datasets made from original datasets. 
 We removed 75 \% ratings from each user (item) to make them sparse dataset. 
 It may be noted that  during sparsing process all ratings of few items (users) are deleted  fully. 
Description of these  sparse datasets is given in Table \ref{t2}.

 \begin{table}[htp]
   \begin{center}
   \begin{tabular}{|c|c|r|r|r|r|r|r|} \hline
   \emph{Dataset } & \emph{Purpose} & $|\mathcal{U}|$  & $|\mathcal{I}|$ &\emph{\#Ratings(RT)} & $|\mathcal{U}|/|\mathcal{I}|$ & $\kappa=\frac{RT\times 100}{|\mathcal{U}| \times|\mathcal{I}|}$  \\
   \hline 
    $Movielens$  & Movie& 6040 & 3517& 0.25 M& 1.71 & 1.18  \\
    $Yahoo$  & Music & 15082 & 1000 & 0.08M & 15.08 & 0.51\\
    $Netflix$  & Movie& 4141 & 8094& 0.25 M & 0.51 & 0.69 \\
   \hline 
   \end{tabular}
   \end{center}
   \caption{Description of the Sparse datasets used in the experiments.} \label{t2}
   \end{table}

\begin{center}  
\begin{figure}
   \includegraphics[width=12cm]{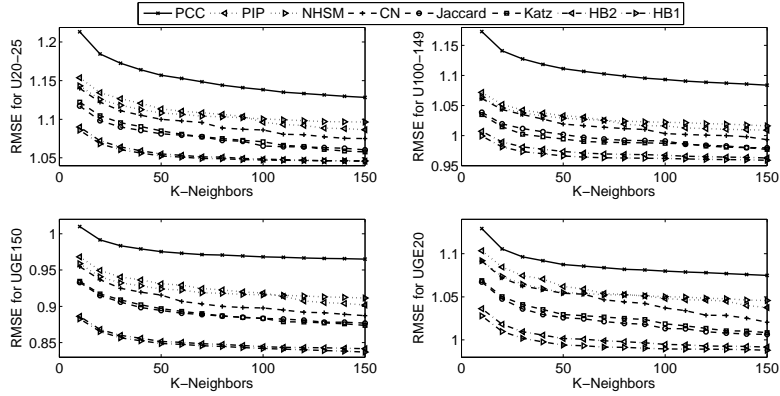}
\caption{RMSE in Movilens }
\label{f2}
\end{figure}
\end{center}

\begin{center}  
\begin{figure}
   \includegraphics[width=12cm]{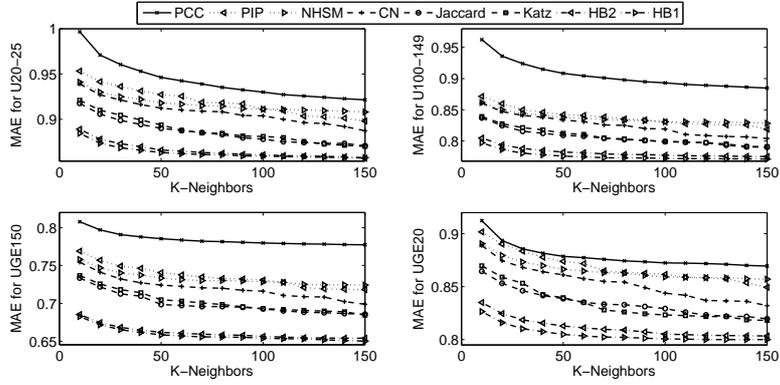}
\caption{MAE in Movilens }
\label{f3}
\end{figure}
\end{center}

\begin{center}  
\begin{figure}
   \includegraphics[width=12cm]{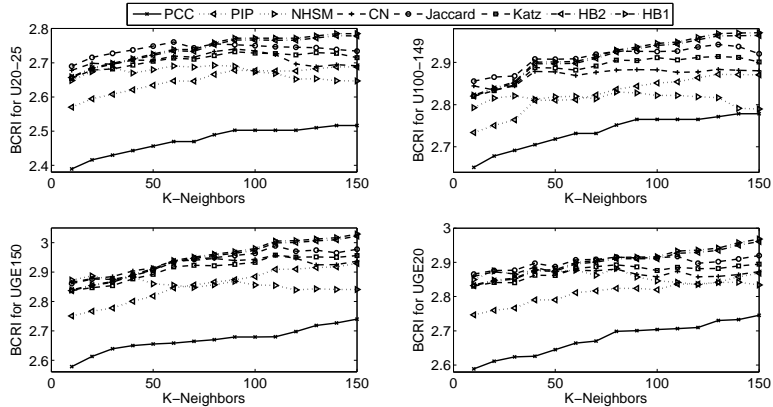}
\caption{BCRI in Movilens }
\label{f4}
\end{figure}
\end{center}

 \subsection{Metrics}
  We use various evaluation metrics to compare the accuracy of the  results obtained by our 
  network based CF and other  neighborhood based CFs. Two popular quantitative  metrics 
  ({\bf Root Mean Squared Error and Mean Absolute Error}) and one popular qualitative 
  metric ({\bf F1 measure}) are used. For the shake of readability, we discuss them briefly. 
  Finally, we introduce a new qualitative measure termed as {\em Best Common Rated Item (BCRI)} 
  to address the drawback of the F1 measue in recommendation scenario. 
  
  \begin{enumerate}
  
  \item 
   \textbf{Root Mean Squared Error(RMSE):} 
   Let $X_{u}$=$[e_{u1},e_{u2},e_{u3}...e_{u1m}]$ be the error vector for  
   $m$ rating prediction of a user  $U_{u}$. A smaller value indicates a better accuracy. 
   Root Mean Square Error(RMSE) for a user is computed as follows.
  \begin{equation*}
  RMSE= \sqrt{\frac{\sum_{i=1}^{m}e_{ui}^{2}}{m} } 
  \end{equation*}

  \item 
   \textbf{Mean Absolute Error(MAE):} 
   Mean Absolute Error measures average absolute error over $m$ predictions 
   for a user $U_u$. It is computed as follows.

  \begin{equation*}
   MAE=\frac{\sum_{i=1}^{m} |e_{ui}|}{m}
   \end{equation*}

 \begin{item}
   \textbf{F1 Measure:} 
   Many recommender systems provide a list of items $L_{r}$ to an active user 
   instead of predicting ratings. There are two popular metrics to evaluate 
   quality of a RS in this scenario: $(i) Precision$, which is the fraction 
   of items in $L_{r}$ that    are relevant and $(ii) Recall$, which is the 
   fraction of total relevant items that are in the recommended list $L_{r}$. 
   A list of relevant items $L_{rev}$ to a user is the set of items on which 
   she made high ratings (i.e, ≥ 4 in MovieLens dataset) in the
   test set. Therefore, $Precision$ and $Recall$ can be written as follow.
  \begin{center} $Precision =
   \frac{| L_{r} ∩ L_{rev} |}
   {| L_{r} |} $and $Recall =
   \frac{| L_{r} ∩ L_{rev} |}
      {| L_{rev} |}$
   \end{center}
   However, there is always a trade-off between these two measures. 
   For instance,   increasing the number of items in $L_{r}$ increases 
   $Recall$ but decreases $Precision$.
   Therefore, we use a measure which combines both called 
   $F1$ measure in our experiments.
   \begin{equation*}
   F1 =\frac{
      2 × Precision × Recall
      }{Precision + Recall}
   \end{equation*}
   
   \end{item}
 
 \item 
 \textbf{Best Common Rated items:} The F1 measure is very  popular among information retrieval 
 community. However, it is not suitable measure in the following scenario. 
 There are users who are very lenient in giving 
 ratings, e.g $Tom$ has given minimum rating value of 4 to items. On the 
 other hand,  there are users who are very strict on giving ratings, 
 e.g $Siddle$ has given at maximum rating value 2 to items. For strict 
 users, it is difficult to find  $L_{rev}$  the set of items on which she made high 
 ratings (i.e, ≥ 4 ) as such ratings are not present. Also if no predicted 
 rating is $\geq 4$, we can not compute $L_{r}$. To measure performances in such 
 scenario, we introduce a new metric termed as Best Common Rated Items (BCRI). It 
 determines whether best actual rated entries are also best predicted 
 entries or not. \\
  Let $BAI=\left \{  ba_{1},ba_{2},ba_{3},ba_{4},ba_{5}\right \}$ be the 
  set of top $t$ best rated items by a user and  $BPI$ be the set of top $t$ best predicted 
  items for the user. $BCR$ is computed as follows.
  \begin{equation*}
  BCRI=BAI \cap BPI
  \end{equation*}
  
 \end{enumerate}

\begin{center}  
 \begin{figure}
    \includegraphics[width=12cm]{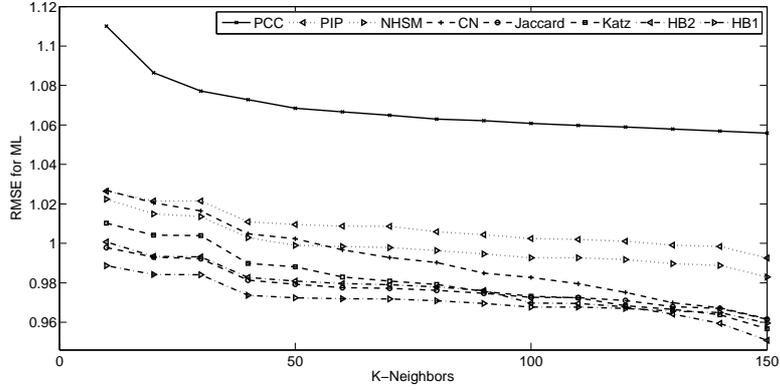}
 \caption{RMSE in Sparse ML }
 \label{f5}
 \end{figure}
 \end{center}
\begin{center}  
 \begin{figure}
    \includegraphics[width=12cm]{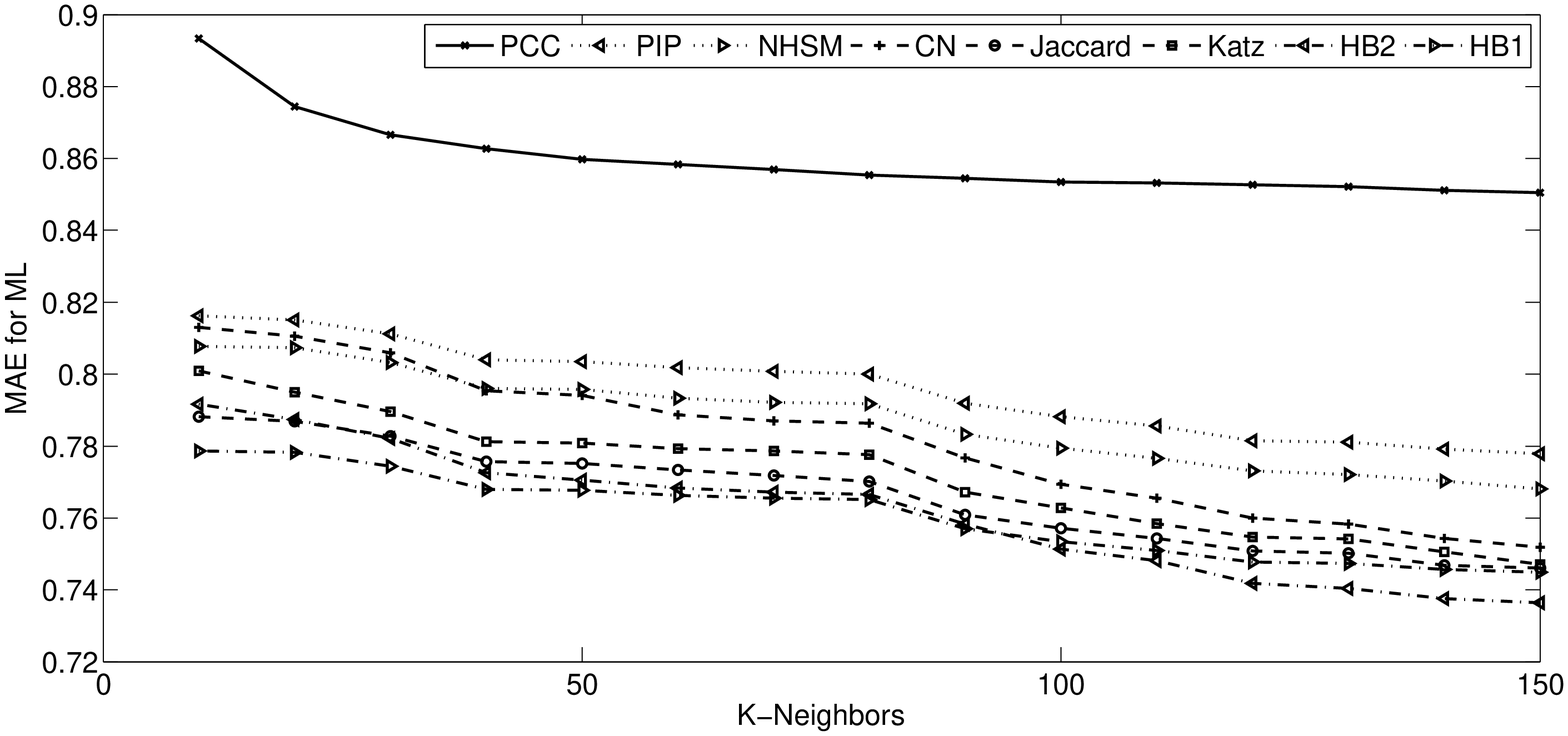}
 \caption{MAE in Sparse ML }
 \label{f6}
 \end{figure}
 \end{center}

 \begin{center}  
  \begin{figure}
     \includegraphics[width=12cm]{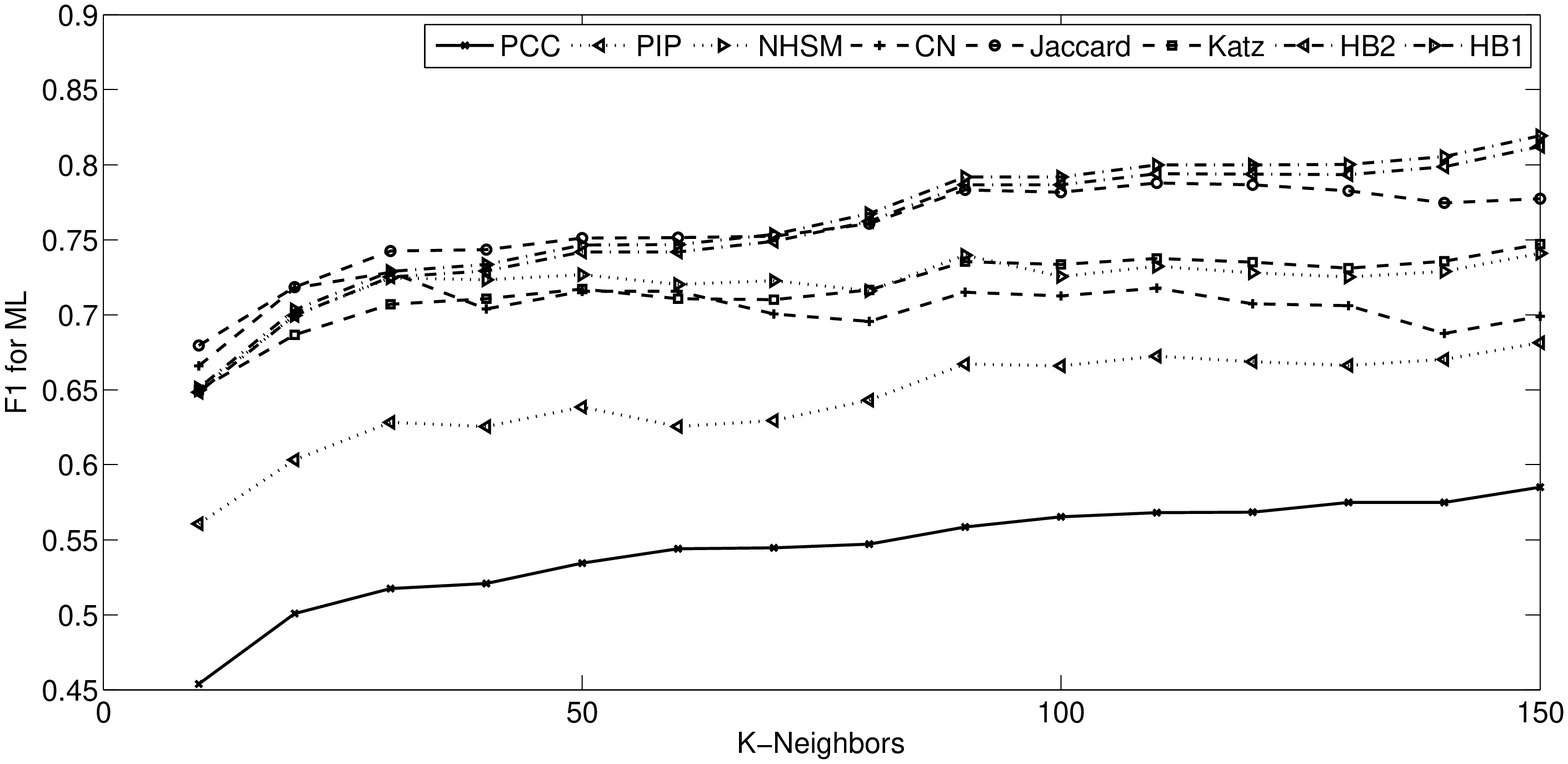}
  \caption{F1 in Sparse ML }
  \label{f7}
  \end{figure}
  \end{center} 
\subsection{Experimental Results and Analysis}

In  experiment setup 1, from each subset we selected users or items and predicted 
deleted ratings using all other users or items in the corresponding rating dataset. 
As we know traditional similarity measures use ratings of only co-rated items in case of 
user-based CF, whereas, they use  ratings of common users in case of item-based CF. 
Therefore,  in many situation we cannot find similarity between a  pair of users (items). 
This is reflected in figure\ref{f1}. In figure \ref{f1}, we get many times {\em NaNs}
during similarity computations specifically while computing similarity for  users in the first 
group \textbf{U20-25}. It can be noted that if  similarity come out to be NaN, 
we can not use it for prediction purpose.  This shows that sparsity is a big issue 
in computing similarity using traditional similarity measures.

For Yahoo dataset we  use item-based method as number of items is quite 
less than number of users. In this dataset,  when we select items from \textbf{IGE20} 
for computing similarity, we find many {\em NaN}s (8460) for 150 items. Similar trends are 
found for other groups of items. 

Similarly in Netflix data set, When we select users from \textbf{UGE20} total 
number  of NaNs during similarity calculation for 150 users is 35816 and when we 
select users from \textbf{U20-25} it is 14535(for 6 users), where as 
for \textbf{U100-149} , \textbf{UGE150}, the numbers are significantly less, {\em i.e} 12881, 2932 respectively .

So we can see that few co-rated entries are hindrance in determining 
similarities between a pair of  user or items. We use structural 
similarity measures to get rid of this problem.

We  begin analyzing results of the experiments with  Movielens dataset. Results of experiment setup 1 
on Movielens is shown in Figure \ref{f2}- \ref{f4}. 
In general RMSE, MAE decrease, 
while BCRI increases with increase of nearest neighbors. The 
state-of-the-art similarity measure NHSM performs better than PIP in RMSE 
(Figure \ref{f2}), MAE (Figure \ref{f3}), 
BCRI. However, for large value of $K$  PIP outperforms NHSM and other traditional measures in RMSE, MAE.  
From Figure \ref{f2}- \ref{f4}, it is found  that structural 
similarities based CFs outperforms PCC  measure based as well as 
state of art similarities based CFs in each metric. 
Hybrid methods outperforms  all other similarity measures in RSME, MAE. However, 
for small value of $K$ Jaccard similarity obtained from network outperforms hybrid methods in BCRI. 
The  \textbf{HB1} is best. Among different categories, prediction performance for users in 
\textbf{UGE150} is  best while  for users in \textbf{UB20-25} is worst, it clearly shows 
that performance is relatively good for dense vectors.

Results of experiment setup 2 on sparse  Movielens subset are  shown in 
figure \ref{f5}-\ref{f7}.   In figure \ref{f5}, it can be noted that traditional similarity 
measure PCC performs poorly badly compared to the state of the art similarity measures. 
Recently proposed NHSM outperforms  PIP measures in RMSE. However, all
structural similarities computed from proposed network of users  outperform
PIP, NHSM and PCC measures in RMSE. Hybrid techniques are found to be 
outperforming other structural similarly measures derived  from the network. 
In Figure \ref{f6}, MAE of the CFs are plotted over increasing value of $K$. Similar trends 
are noted here also. Structural similarly measures based CFs outperform PCC and 
NHSM and PIP measures. Two  hybrid techniques which are proposed to remove 
$K$-neighbor problem is found to be better than other structural similarity measures. 
The first hybrid approach makes least MAE as low as 0.82 at the value of $K=150.$ 
 
The plot in figure~\ref{f7} shows the efficiency of our structural similarity (extracted from network) 
based CFs over others measures based CFs. The F1 measure is used to show the capability of an approach to 
retrieve relevant items in a user's recommended list. It is found that hybrid approach including other structural 
similarly outperform PCC, PIP and NHSM measures  in F1 measures. This facts justifies our claim that 
hybrid approaches with network can address the problem of data sparsity. It can be noted that sparsity 
of the ML subset is $98.82\%$ (Table~\ref{t2}).

\begin{center}  
\begin{figure}
   \includegraphics[width=12cm]{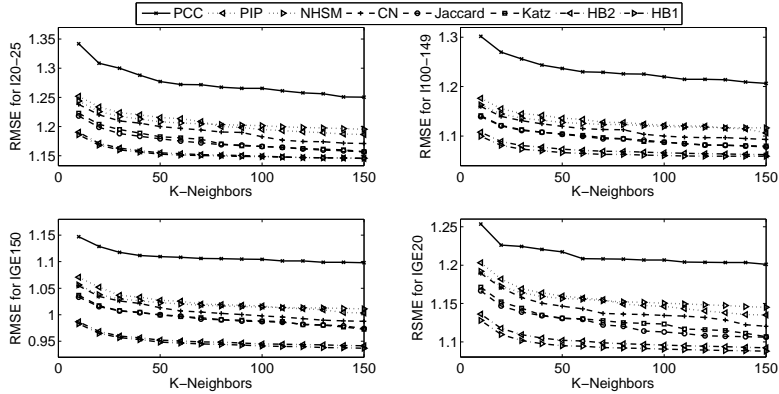}
\caption{RMSE in Yahoo }
\label{f8}
\end{figure}
\end{center}

\begin{center}  
\begin{figure}
   \includegraphics[width=12cm]{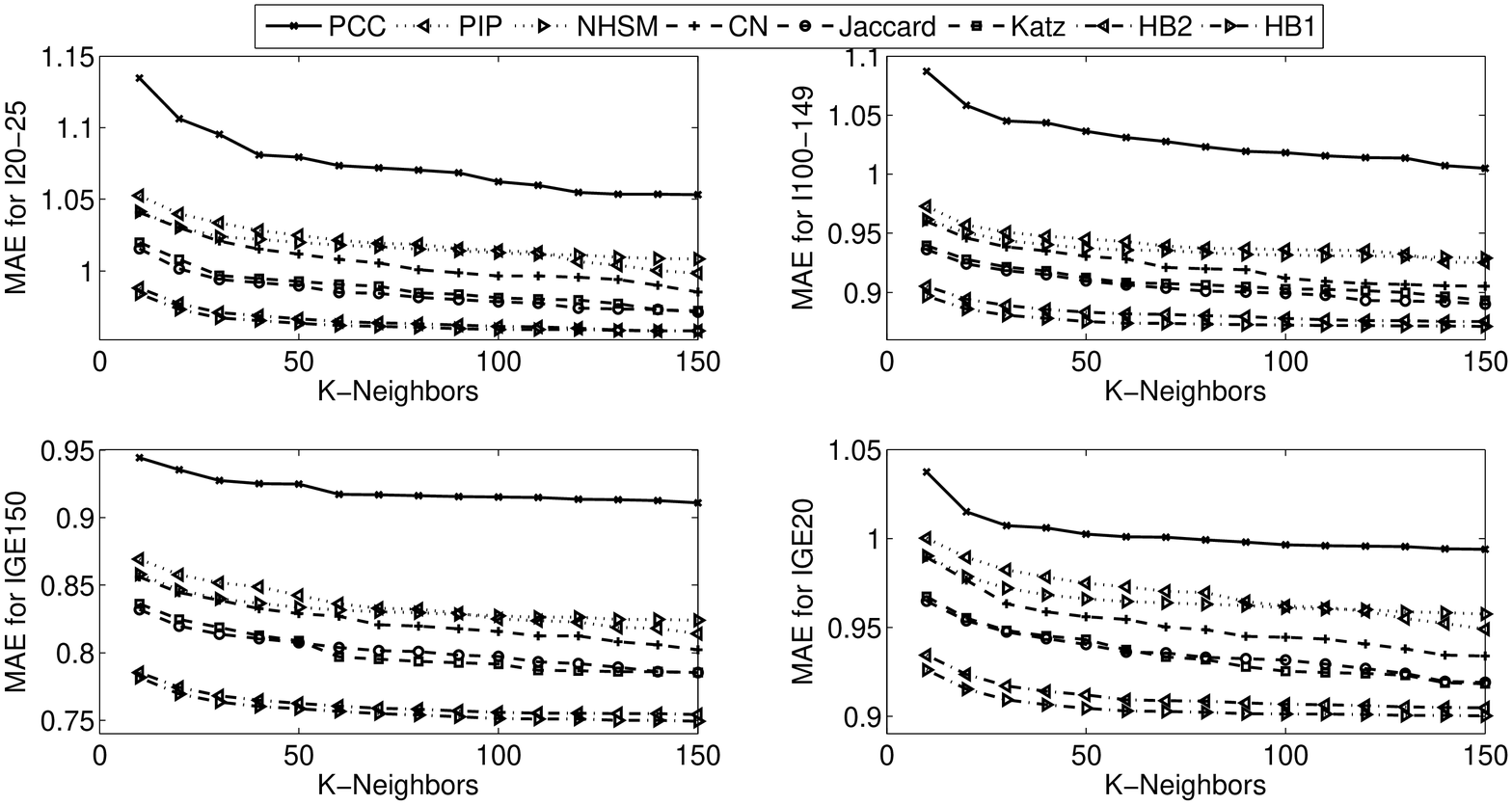}
\caption{MAE in Yahoo }
\label{f9}
\end{figure}
\end{center}

\begin{center}  
\begin{figure}
   \includegraphics[width=12cm]{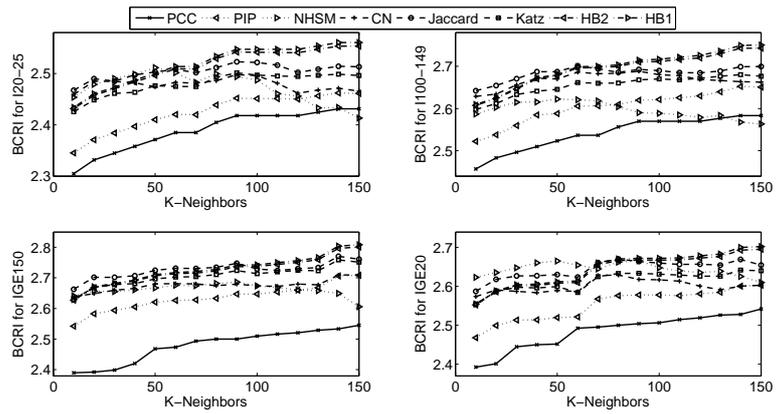}
\caption{BCRI in Yahoo }
\label{f10}
\end{figure}
\end{center}

Results of experiment setup 1 on Yahoo dataset is shown in figure \ref{f8}- \ref{f10}. 
In general RMSE, MAE decreases, while BCRI increases with $K$-$Neighbors$. NHSM 
performs better than PIP in RMSE, MAE, BCRI but for large value of  $K$, PIP performs better 
in RMSE, MAE than NHSM.  It is observed that  structural similarities outperforms PCC and 
state of art similarities in RMSE, MAE. Hybrid 
methods along with other structural measures outperform 
PIP, NHSM, PCC similarity measures in RMSE, MAE and BCRI metrics. Among the proposed measures, 
 \textbf{HB1} is best  in all metrics. It can be noted that we applied item-based CF on Yahoo dataset. 

\begin{center}  
\begin{figure}
   \includegraphics[width=12cm]{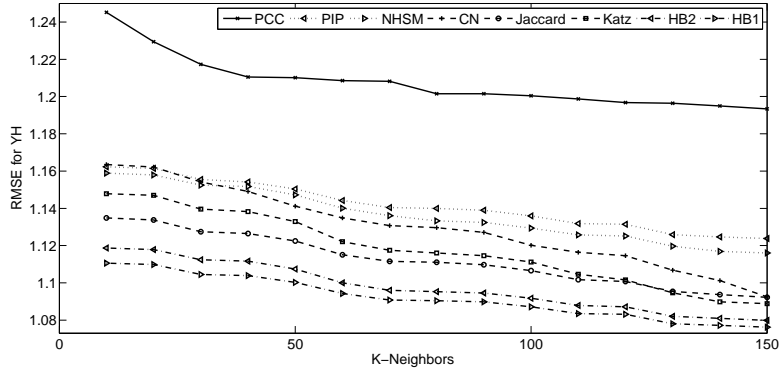}
\caption{RMSE in Sparse YH }
\label{f11}
\end{figure}
\end{center}

\begin{center}  
\begin{figure}
   \includegraphics[width=12cm]{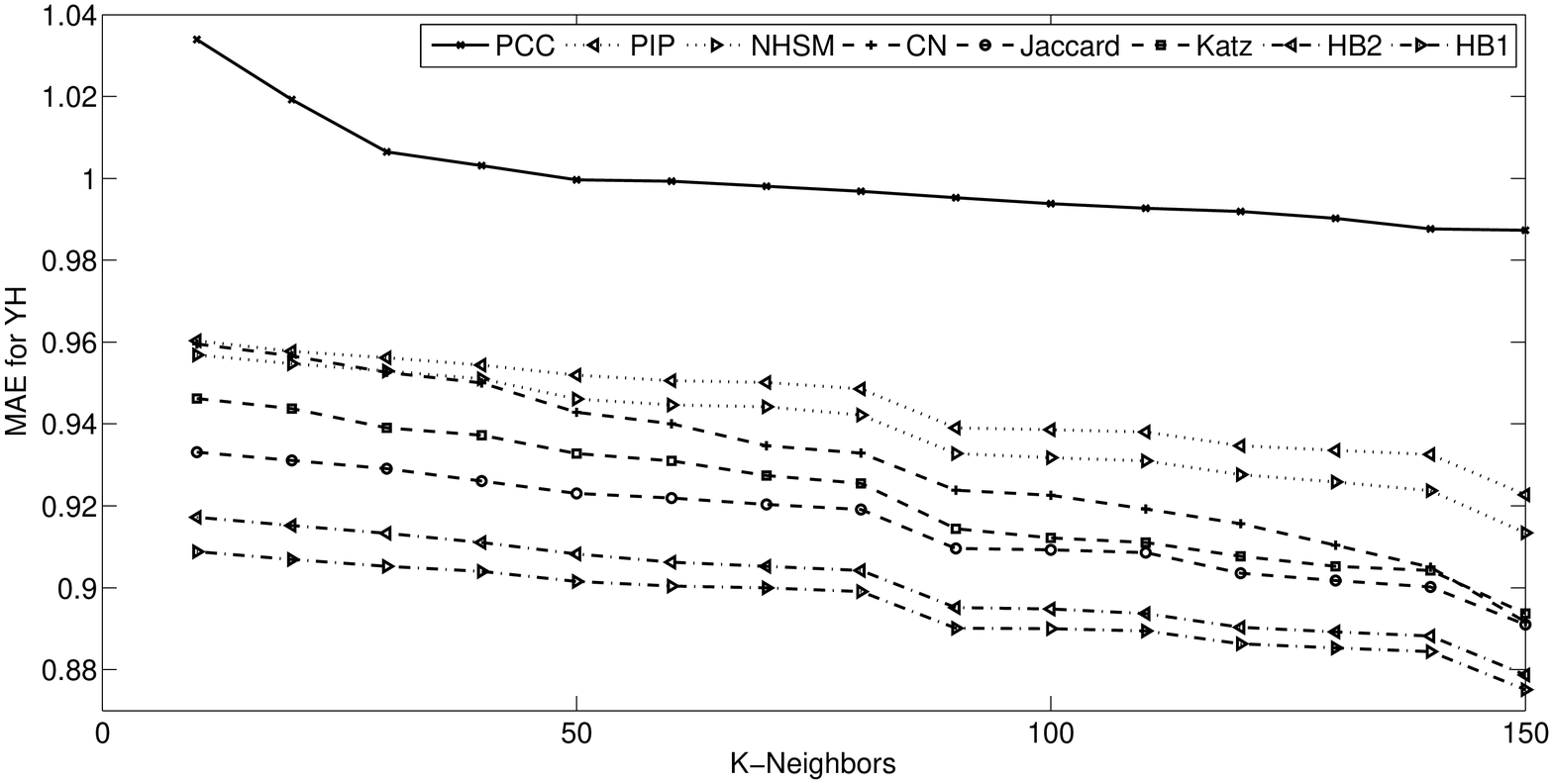}
\caption{MAE in Sparse YH }
\label{f12}
\end{figure}
\end{center}

\begin{center}  
\begin{figure}
   \includegraphics[width=12cm]{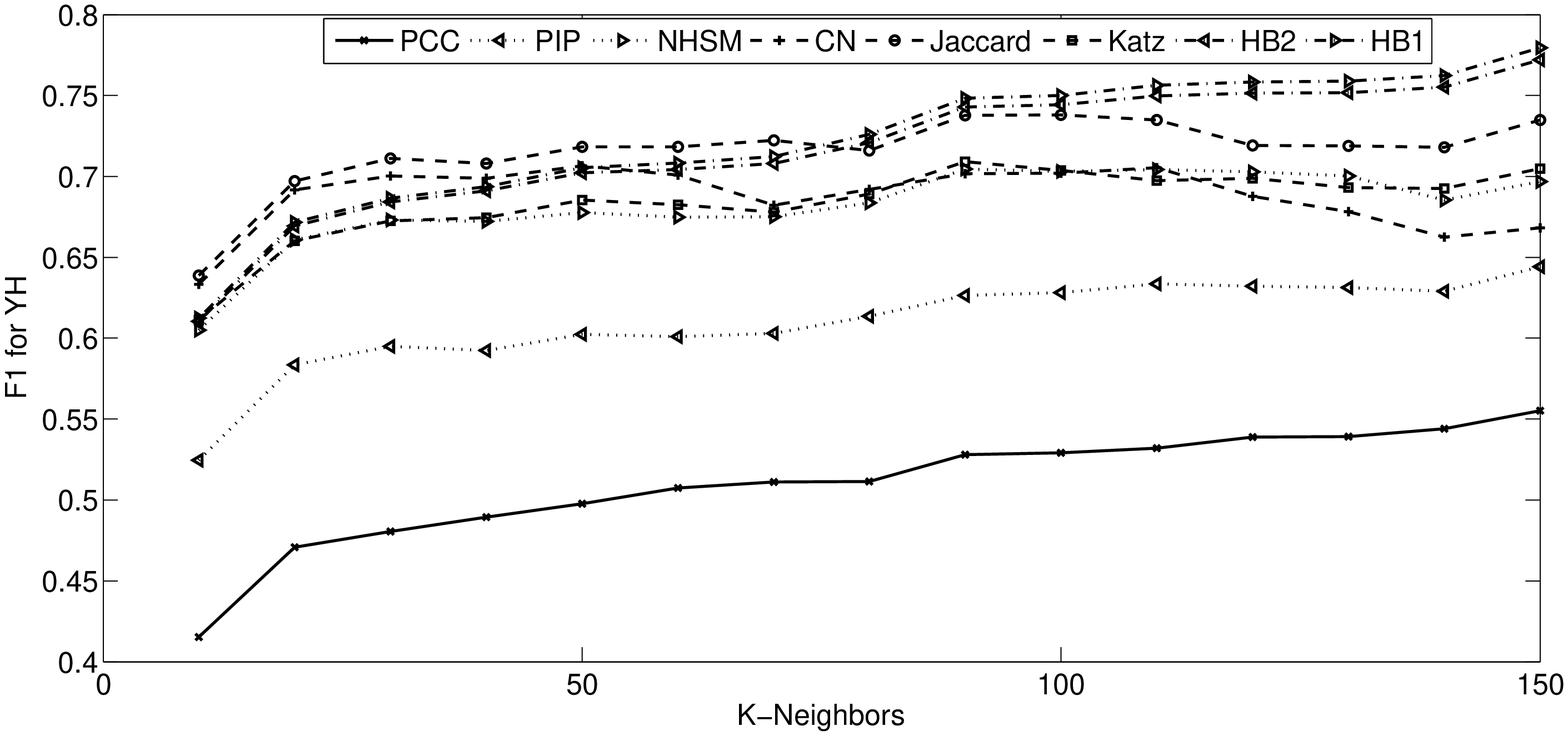}
\caption{F1 in Sparse YH }
\label{f13}
\end{figure}
\end{center}

To show the efficiency of the proposed network approach, we compare the performance of item-based 
CFs on  Yahoo dataset and  results are  reported in figure \ref{f11}- \ref{f13}. In Figure \ref{f11} and 
Figure~\ref{f12}, predictive metrics RMSE and MAE of different similarity measures based CFs 
are plotted over the value of $K$.  In the both plots, it is found that structural similarity 
obtained from item network can provide better RMSE and MAE values compared to PCC, PIP and recently 
introduced NHSM measure. 
In figure~\ref{f13}, F1 measures of different structural similarity based CFs, PCC based CF and 
PIP and NHSM based  CFs are shown in increasing value of $K$. PCC based CF is worst performer 
in F1 measure. The PIP measure based CF outperforms NHSM measure. All similarity extracted 
from item-item network are found to be  outperforming traditional PCC, PIP and recently introduced 
NHSM measure. The hybrid techniques are found to be suitable in this highly sparse dataset. 
It can be noted that network approach is equally successful in item-based CF.

\begin{center}  
\begin{figure}
   \includegraphics[width=12cm]{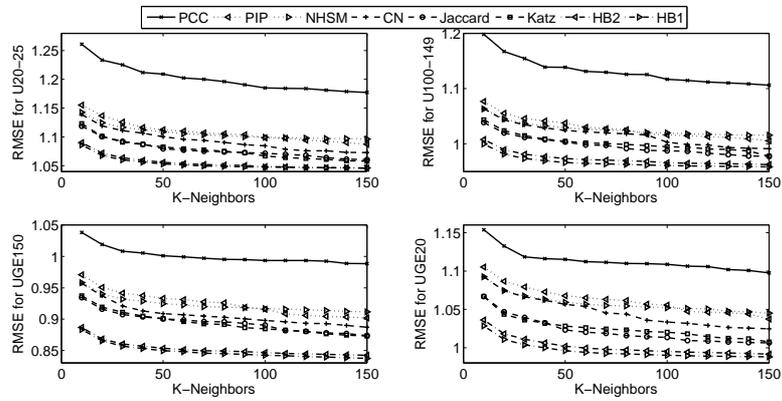}
\caption{RMSE in Netflix }
\label{f14}
\end{figure}
\end{center}

\begin{center}  
\begin{figure}
   \includegraphics[width=12cm]{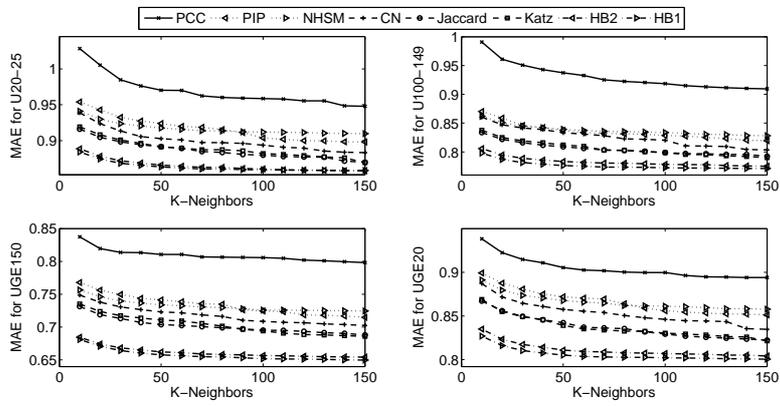}
\caption{MAE in Netflix }
\label{f15}
\end{figure}
\end{center}

\begin{center}  
\begin{figure}
   \includegraphics[width=12cm]{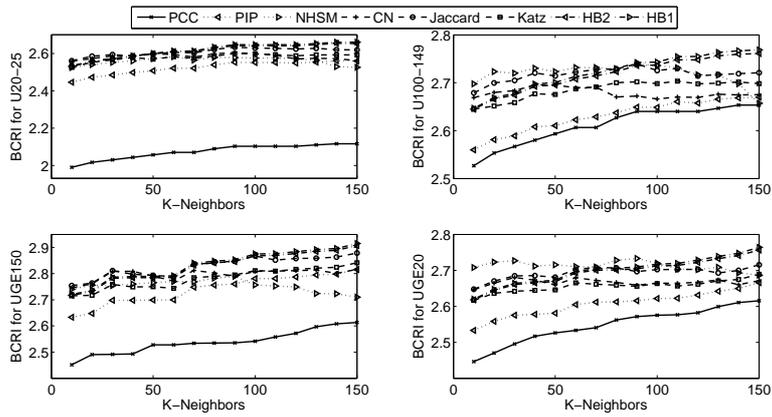}
\caption{BCRI in Netflix }
\label{f16}
\end{figure}
\end{center}

\begin{center}  
\begin{figure}
   \includegraphics[width=12cm]{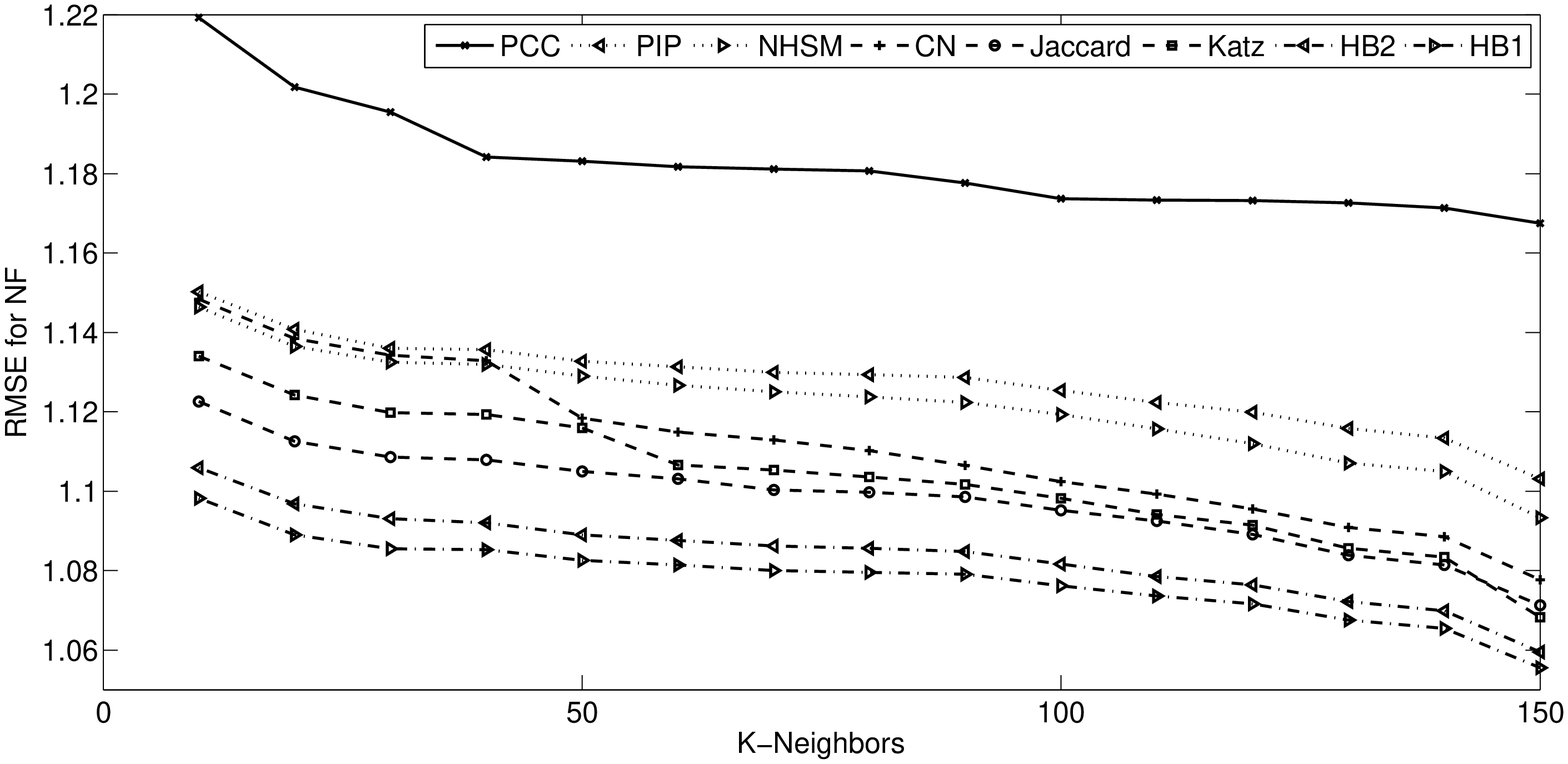}
\caption{RMSE in Sparse NF }
\label{f17}
\end{figure}
\end{center}

\begin{center}  
\begin{figure}
   \includegraphics[width=12cm]{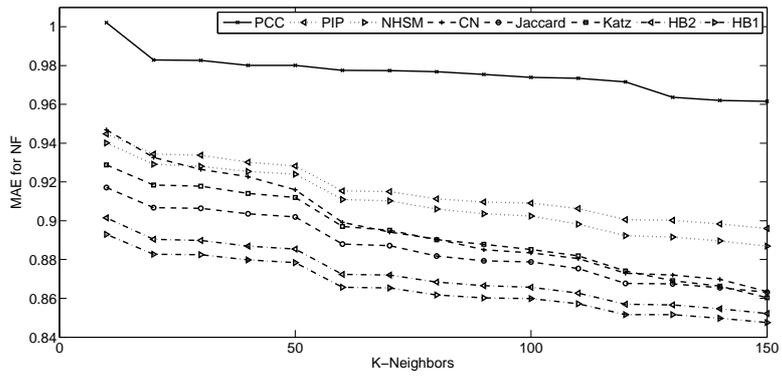}
\caption{MAE in Sparse NF}
\label{f18}
\end{figure}
\end{center}

\begin{center}  
\begin{figure}
   \includegraphics[width=12cm]{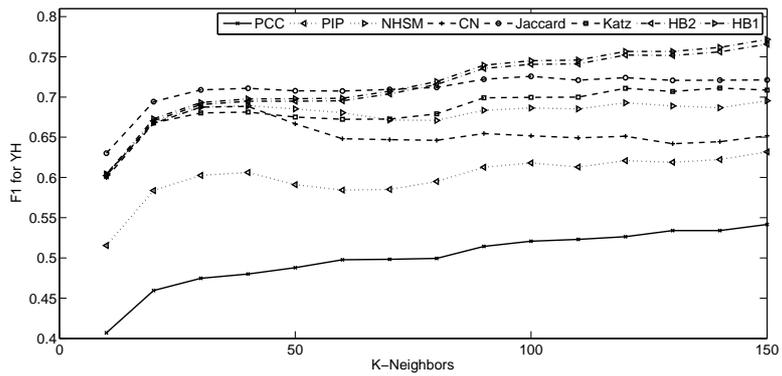}
\caption{F1 in Sparse NF }
\label{f19}
\end{figure}
\end{center}

Results of experiment setup 1 on Netflix is shown in figure \ref{f14}- \ref{f16}. 
NHSM performs better than PIP for smaller values of  $K$ in RMSE, MAE, BCRI  metrics. 
However,  PIP outperforms NHSM in large value of $K$. 
We see that structural similarities outperform PCC and 
state of art similarities in RMSE,  MAE and BCRI measures.  
Hybrid methods outperforms all other similarity measures in RSME, MAE and BCRI metrics. 
The \textbf{HB1} is best  in all metrics.

Experimental results  of experiment setup 2 on sparse Netflix dataset is shown in 
figure \ref{f17}- \ref{f19}. Simialr trends are noted. Structural similarities based CF outperform
PCC, PIP and NHSM based CFs in MAE, RMSE and F1 measure on 
highly sparse Netflix subset. For smaller value of  $K$, NHSM 
performs better than CN. However, with  suitable number of nearest neighbors, 
hybrid methods perform better than other 
similarity measures in RMSE, MAE and $F1$  metrics. The {\bf HB1} is the best in all metrics 
even on sparse datasets.

\section{Conclusion }\label{sec:con}

In this work, we proposed a new outlook to deal with the problem of collaborative recommendation by gainfully using the concept of structural similarity of nodes in a complex network after generating user-user and item-item networks based on the given data. We showed that, the curse of sparsity and K-neighbors problem can be delicately handled in this approach. Thus, we proposed CFs based on structural similarity measures of the user-user and item-item networks individually. Moreover, we introduced two methods which we call hybrid methods using both user-user and item-item networks for $CF$. We verified the effectiveness of these measures by comparing its performances with that of neighborhood based CFs using state-of-the-art similarity measures when applied to a set of real data. The comparison results established that the proposed measures based CFs and hybrid methods outperform existing similarity measures based CFs in various evaluation metrics.

\bibliographystyle{plain}
\bibliography{a}

\end{document}